# Exploring the interplay between Planetary Boundaries and Sustainable Development Goals using Large Language Models

**Authors**

Lamyae Rhomrasi[a], Pilar Manchón[b], Ricardo Vinuesa[c], Francesco Fuso-Nerini[d,e]
J. Alberto Conejero[a], Javier García-Martínez[f], and Sergio Hoyas[a]

**Affiliations**

a: Instituto Universitario de Matemática Pura y Aplicada, Universitat Politècnica de València, 46022 Valencia, Spain
b: Google Research, Mountain View, California, United States
c: Department of Aerospace Engineering, University of Michigan
d: Division of Energy Systems, School of Industrial Engineering and Management, KTH Royal Institute of Technology, Stockholm, Sweden
e: KTH Climate Action Centre, KTH Royal Institute of Technology, Stockholm, Sweden
f: Molecular Nanotechnology Lab, Department of Inorganic Chemistry, University of Alicante, Alicante, Spain

**Corresponding author**

Sergio Hoyas, e-mail: serhocal@mot.upv.es, tel: +34620945962

Instituto Universitario de Matemática Pura y Aplicada, Universitat Politècnica de València, 46022 Valencia, Spain

**Abstract**

By analyzing 40,037 climate articles using Large Language Models (LLMs), we identified interactions between Planetary Boundaries (PBs) and Sustainable Development Goals (SDGs). An automated reasoner distinguished true trade-offs (SDG progress harming PBs) and synergies (mutual reinforcement) from double positives and negatives (shared drivers). Results show 21.1% true trade-offs, 28.3% synergies, and 19.5% neutral interactions, with the remainder being double positive or negative. Key findings include conflicts between land-use goals (SDG2/SDG6) and land system boundaries (PB6), together with the underrepresentation of social SDGs in the climate literature. Our study highlights the need for integrated policies that align development goals with planetary limits to reduce systemic conflicts. We propose three steps: (1) integrated socio-ecological metrics, (2) governance ensuring that SDG progress respects Earth system limits, and (3) equity measures protecting marginalized groups from boundary compliance costs.

---

The Planetary Boundary (PBs) framework [1] defines the limits of human activities required to maintain the stability of a stable and resilient Earth [2]. They establish a "safe operating space" for humanity, beyond which the risk of generating abrupt or irreversible large-scale environmental changes increases. The nine identified boundaries encompass key Earth system processes, such as climate change and biogeochemical flows, emphasizing the interconnections between these critical processes and the need for a holistic approach to sustainability.

Since its introduction, the PB framework has been widely used to assess the impact of various research fields on the Earth system. For instance, the Rockefeller Foundation–Lancet Commission on Planetary Health [3] employs the PB concept to explore the intricate relationship between human health and the natural systems of the planet. Their research identifies several critical environmental changes that pose risks to human well-being, including climate change, ocean acidification, land degradation, water scarcity, overfishing, and loss of biodiversity.

On the other hand, the Sustainable Development Goals (SDGs), established by the United Nations in 2015, consist of 17 global objectives designed to eradicate poverty, protect the planet, and foster prosperity for all [4]. These goals were set to be achieved by 2030, but it has become increasingly evident that many of the SDGs will not be met.

In a recent study, Vinuesa et al. [5] examined the role of artificial intelligence in advancing SDGs, highlighting the need for regulatory guidance. Their work inspired the present study, which builds on the idea that developing an integrated sustainability agenda, one that effectively addresses both social and environmental priorities, requires a deep understanding of the complex interplay between PBs and the SDGs.

**Methodology and Results**

Data were extracted from 40,037 open-access climate articles using five sequential LLM queries per document [6, 7]. The corpus analyzed consists of general climate and sustainability science articles, which may simultaneously reflect several PBs and SDGs, even if the original authors did not explicitly frame their work in those terms. We did not perform screening of the retrieved records, since our aim is to study the entire landscape rather than pre-filtering, in order to capture the full range of implicit and explicit PB–SDG linkages. Importantly, we used the OpenAlex database because it allows large-scale access to full-text records of open-access publications, unlike Web of Science or Scopus, which are closed-access and do not permit full-text integration at this scale. In our pipeline, the LLMs were applied to the full text of each article (not only the abstracts), which is a crucial advantage of using OpenAlex for this type of large-scale analysis.

These articles were primarily obtained in PDF format, often as author-copy versions. To extract structured textual data, we utilized the AI-based library Grobid, which converted the documents into XML/TEI format. This process systematically removed elements irrelevant to the present study, such as figures, acknowledgments, and bibliographic sections, ensuring a clean and well-structured dataset for subsequent analysis.

For this study, we used Google's Gemini-1.5 Flash. A key advantage of this model is its extensive one-million-token context window, which enables the analysis of lengthy scientific articles in a single prompt. To mitigate potential biases from simple keyword matching, the model's knowledge was supplemented with explicit definitions of PBs and SDGs.

The methodology involves sequentially applying five distinct queries to the LLM, with prompts detailed in subsequent sections. The processing pipeline for each individual article follows these steps:

- **SDG Allocation**: Assigns Sustainable Development Goals to the article without predefined constraints.
- **PB Allocation**: Identifies the relevant Planetary Boundaries for the article, also without constraints.
- **SDG–PB Relationship Classification**: Determines whether the interaction between each SDG–PB pair represents a synergy, a trade-off, or a neutral link.
- **SDG–PB Causality Analysis**: Establishes the directionality of the relationship, i.e., whether an SDG influences a PB or vice versa.

Despite the model's large context window, we imposed a limit of 20 SDG–PB pairs per query in Steps 3 and 4. Although a higher number is technically feasible, performance degradation occurs due to output token limitations, which adversely affects the quality of reasoning and results in less robust justifications. To ensure reliable identification of synergies and trade-offs, we required clear textual evidence of measurable actions, policies, or outcomes that directly demonstrate a relationship between an SDG and a PB. Additionally, the prompt explicitly discouraged speculative or overly broad assumptions, prioritizing evidence-based reasoning to strengthen the validity of the findings.

Finally, to enhance the classification of synergies and trade-offs, an additional fifth prompt was introduced to validate whether the identified interactions were genuinely reflected in the text, using an experimental reasoning model (Gemini 2.0 Flash Thinking). In this step, synergy classifications were further split into three groups: "Generality," "Misled by Positivity," and "Actual Synergy." This refinement was essential for distinguishing actual synergies—where progress in one goal directly benefits another— from generalized statements and overly optimistic predictions shaped by the positive tone of much scientific literature. Similarly, trade-offs were classified into three groups: "Actual Trade-off," "Generic Negative Association," and "Double Negative (Co-Degradation)," to differentiate between genuine trade-offs (where progress in one domain pressures another), cases of broad negative associations lacking textual evidence, and co-degradations (where harm to one domain negatively affects another). This clarification strengthened the robustness of the analysis and allowed a deeper understanding of the underlying mechanisms.

It should also be noted that an analysis at the SDG target level, while certainly desirable, would be prohibitively complex given the large number of targets (169), which would multiply the classification effort; hence, this study focuses on the goal level, consistent with other large-scale analyses. Regarding the reliability of the LLM classifications, our method has been previously validated in [8], which substantially reduces the risk of hallucinations and ensures that the automated reasoning remains within the scope of the textual evidence. Further details on the methodology are provided in the SI.

**Results and Discussion**

**Figure 1** summarizes the quantitative SDG–PB interactions. The main panel features 17 plots, one per SDG, each with nine bars representing PBs. Labels show link counts, and stacked areas within bars depict the proportions of synergies (green), neutral (yellow), and trade-offs (red). Initial analysis showed 33.8% synergies, 44.9% trade-offs, and 19.5% neutral interactions overall. Our Reasoner refined this (darker overlays), identifying 28.3% as validated TS (mutual reinforcement) and 21.1% as TT (SDG progress increases the PB pressure).

Figure 1 highlights the disproportionate trade-offs between PB2 and SDG14, with 75.5% of PB2–SDG14 links classified as trade-offs—1.6 times the global average. Given PB2's low presence (12%) and SDG14's moderate coverage (21%), this contrast is interesting. Since improving SDG14 should not worsen PB2, and vice versa, the reasoner identified 97.5% of these cases as DN, where ocean acidification and marine ecosystem decline are driven by the same external pressures. This highlights a critical policy insight: interventions must address the root causes, particularly anthropogenic $CO_2$ emissions, which degrade marine ecosystems (SDG14) and breach ocean acidity boundaries (PB2).

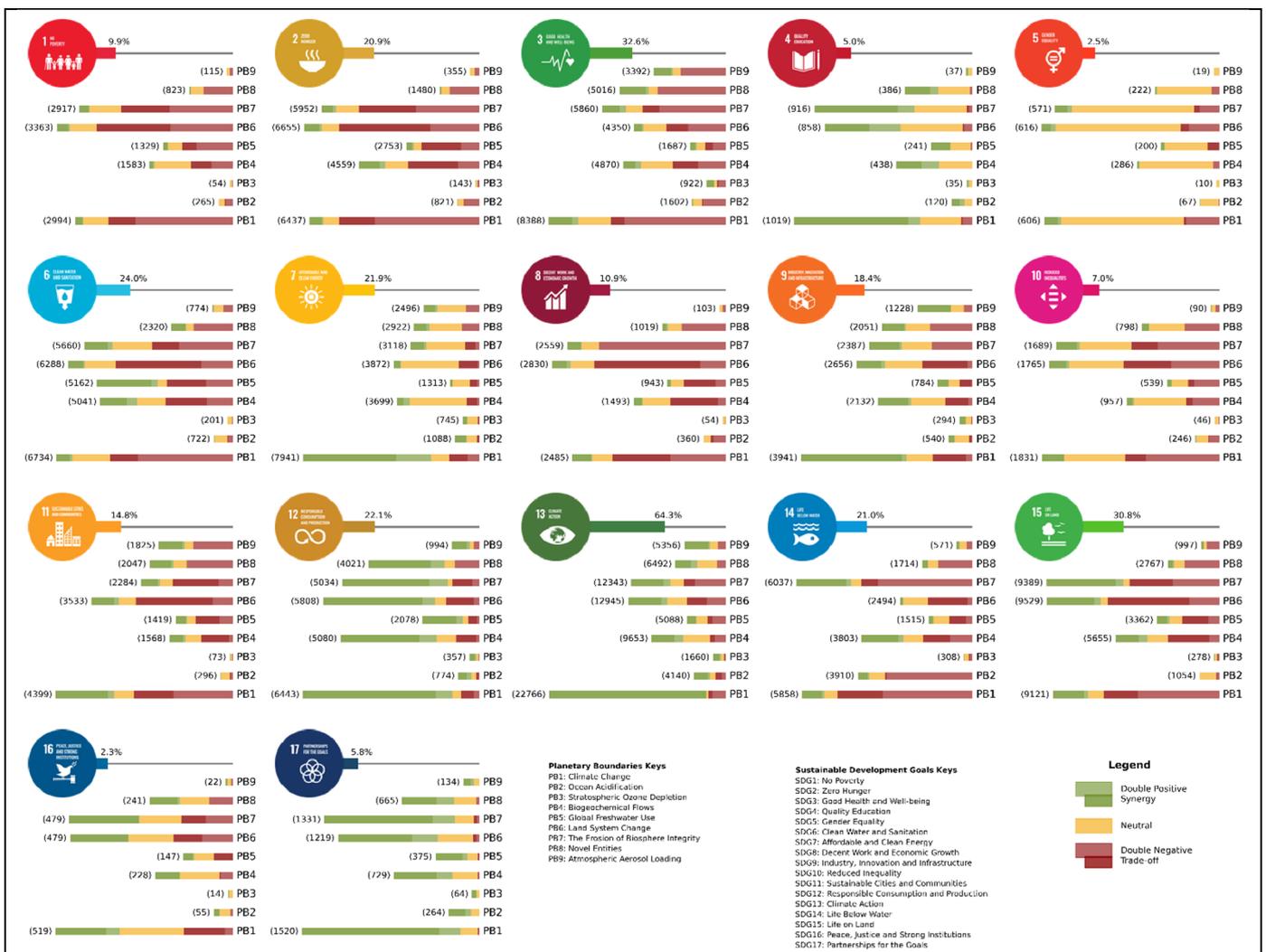

**Figure 1.** Connections between the SDGs and PBs. Each subpanel shows the interactions between one of the 17 Sustainable Development Goals and the nine Planetary Boundaries. For each SDG, the top bar indicates the proportion of papers in the database related to that SDG. The stacked bars below represent the proportions of synergies (green), neutral interactions (beige), and trade-offs (red) for each PB. Darker overlays within these bars denote the proportions of validated "true" synergies (TS) or trade-offs (TT), distinguishing them from double positives or double negatives (lighter overlays). Bars are normalized by the maximum number of interactions per SDG, and the total number of interactions is indicated to the left of each bar.

Focusing on SDG14 or PB2 in isolation, without tackling shared drivers such as climate change, will likely prove insufficient for achieving sustainability in the marine environment.

Second, Land System Change (PB6), present in 42.1% of the documents, showed frequent conflict, with 52.5% of its SDG interactions involving trade-offs (including DN), versus 25.3% synergies and 19.4% neutral interactions. Notably, SDG12 (Responsible Consumption and Production) deviated strongly, exhibiting 69.2% synergy (88.8% TS), indicating that sustainable production may mitigate land degradation. For instance, regulated tourism and eco-certified infrastructure in Kazakhstan have reduced soil erosion and habitat fragmentation [9], a finding correctly identified by our LLM.

Our data revealed frequent conflicts between ensuring food security (SDG2 – Zero Hunger), clean water (SDG6 – Clean Water and Sanitation), and preserving land systems (PB6), with trade-offs identified in 78.1% and 70.0% of cases, respectively. This reflects the pressure on land quality from agricultural expansion and water demand, complicating the balance between resource use and sustainability. Examples include maize expansion impacting land cover [10] and EU biomass imports driving deforestation [11].

Despite thematic links, SDG15 (Life on Land) registered 62.4% trade-offs with PB6 (73.3% TT), indicating conflicts between habitat protection and extractive land use, such as altered ecosystems from adaptive forest management [12]. SDG13 (Climate Action) showed context-dependent effects on PB6, with nearly equal synergies (38.5%, 84.3% TS) and trade-offs (38.0%, comprising 50.2% TT and 24.1% DN). The LLM flags consistent trade-offs from biofuel policies, which can displace crops and ecosystems [13].

Third, environment- and economy-related SDGs dominate interactions, while social goals such as SDG16 (Peace, Justice and Strong Institutions, 2.3%), SDG5 (Gender Equality, 2.5%), and SDG4 (Quality Education, 4.0%) are underrepresented in the climate literature, yet show 40% trade-offs when linked. Economic activities often framed as poverty reduction (e.g., agricultural intensification, resource extraction, and urbanization) can exacerbate PB pressures and disproportionately harm marginalized groups. Examples include resource extraction depleting reserves and displacing communities under weak governance [14], and urbanization driving biosphere degradation alongside social inequalities [15].

Directionality analysis revealed that 69.4% of the interactions were driven by PB-to-SDG pressures. However, industrial and energy SDGs (7, 9, 12) predominantly drive PB impacts, likely due to their focus on proactive interventions, whereas other SDGs respond to environmental pressures. For instance, SDG7's biofuel expansion can increase land pressure, SDG9's infrastructure may alter land and water systems, and SDG12's initiatives could unintentionally overuse resources. This suggests a potential divide between impact-driving and mitigating SDGs, although interactions are highly context-dependent and require full dataset analysis for a comprehensive understanding.

**Conclusions**

We analyzed SDG–PB synergies and trade-offs using **Gemini 1.5 Flash**, with a reasoner differentiating true trade-offs (improving one worsens another) from double negatives (both decline together) and validating synergies. This yielded 21.1% true trade-offs,

28.3% synergies, and 19.5% neutral interactions, with the remainder being double synergies and trade-offs. Ocean Acidification (PB2) and Life Below Water (SDG14) are frequently presented as double negatives, declining together due to shared external pressures rather than reinforcing each other synergistically. This underscores a critical policy need: tackling the root causes, particularly anthropogenic $CO_2$ emissions, is essential for simultaneously improving marine ecosystem health and respecting ocean acidity limits.

Land System Change (PB6) conflicts with Zero Hunger (SDG2) and Clean Water and Sanitation (SDG6), showing 78.1% and 70.0% trade-offs, respectively, reflecting the resource strain of food production. Additionally, climate literature underrepresents social goals such as SDG16 (Peace, Justice, and Strong Institutions), SDG5 (Gender Equality), and SDG4 (Quality Education), with 40% of their interactions classified as trade-offs. This suggests a bias toward environmental and economic dimensions, particularly from a Global North perspective, potentially marginalizing SDGs tied to equity and human vulnerability.

National planning must integrate PB–SDG feedback to avoid damaging trade-offs, such as land system (PB6) conflicts with food and water goals (SDG2/SDG6), by predicting resource pressures. We should prioritize underrepresented social goals, often negatively framed in climate studies, to enhance equity and build resilience for environmental action. Techno-optimism, such as reliance on biofuels, must be critically evaluated: while aiming to advance climate action (SDG13), biofuels can exacerbate land pressure (PB6) and undermine sustainability.

To reduce these trade-offs, we propose three steps:

- **Use integrated metrics**, such as dashboards showing resource footprints alongside poverty and inequality, to reveal socio-ecological links.
- **Strengthen governance**, for instance, through scientific councils setting PB-based standards or budgets that guide SDG actions.
- **Ensure equity**, for example, by allocating environmental tax revenues to Just Transition programs that support communities affected by boundary compliance

**Acknowledgements**

SH acknowledges the funding of project **PID2021-128676OB-I00**, and JGM the project **PID2021-1287610B-C21**, both funded by **MCIN/AEI/10.13039/501100011033** and by *"ERDF A Way of Making Europe"*, European Union.

JGM and JAC are also grateful to the **Generalitat Valenciana** program *Prometeo* through **CIPROM/2023/2** and **CIPROM/2022/21**.

RV and FFN acknowledge the funding provided by **Digital Futures**, within their Demonstrator Project program.